\documentstyle[aps,prl,multicol]{revtex}

\begin{document}

\title{Random Quantum Spin Chains: A Real-Space Renormalization Group Study}

\author{E. Westerberg, A. Furusaki,\cite{AF} M. Sigrist, and P. A. Lee}

\address{Department of Physics, Massachusetts Institute of Technology,
Cambridge, Massachusetts 02139}

\maketitle

\begin{abstract}
Quantum Heisenberg spin chains with random couplings and
spin sizes are studied using a real-space renormalization group technique.
These systems belong to a new universality class of disordered
quantum spin systems realized in {\it e.g.}
${\rm Sr}_3{\rm CuPt}_{1-x}{\rm Ir}_x {\rm O}_6$.
The low-energy fixed point is characterized by the
formation of weakly coupled large effective spins.
At low temperature $T$ the entropy obeys a power law $T^\delta$
($\delta\approx0.44$), and the susceptibility follows a Curie-like behavior.
\end{abstract}

\begin{multicols}{2}
\narrowtext

A class of quasi-one-dimensional (1D) spin systems with an interesting
type of disorder was recently discovered in the compound
${\rm Sr}_3{\rm CuPt}_{1-x} {\rm Ir}_x {\rm O}_6$.
In the pure material, the Cu (spin $S=\frac{1}{2}$)
and Pt ($S=0$) alternate along chains and the
Cu spins interact antiferromagnetically with each other \cite{wilk}.
In the compound ${\rm Sr}_3{\rm CuPt}_{1-x}{\rm Ir}_x {\rm O}_6$
a fraction of the spinless Pt atoms is replaced by Ir with a
spin $S=\frac{1}{2}$ that couples ferromagnetically
to the neighboring Cu spins.
Thus this alloy represents a quasi-1D system with random ferromagnetic
(FM) and antiferromagnetic (AF) bonds of fixed strength
\cite{footnote2}.
Measurements of the uniform susceptibility as a function of temperature
($2{\rm K}<T<300{\rm K}$) revealed a crossover around $10{\rm K}$
between the standard high-temperature and a low-temperature
Curie behavior \cite{nguy}.

In a recent theoretical study we considered a related model of a quantum spin
chain with random FM and AF nearest-neighbor coupling of equal strength
$J$ \cite{furu}. The analysis of high-temperature series for the susceptibility
compares qualitatively well with the experimental data~\cite{footnote4}.
Our analysis of the entropy led to the surprising result that a considerable
fraction of the degrees of freedom remains uncorrelated down to a
temperature $k_BT\sim J/5 $.
This unusual property together with the Curie-like behavior of
the susceptibility indicated the existence of very weakly coupled
effective spin degrees of freedom in the low-energy regime,
which we identify in the following way.
Our system consists of a sequence of alternating FM and AF segments
of random length.
At an intermediate energy scale ($\sim J$) the spins within each segment
lock into their local ground state with a variety of spin values
corresponding to the smallest or largest possible
total spin for AF or FM segments, respectively.
Because of the quantum nature of the spins these effective degrees of
freedom couple rather weakly and behave like independent spins for
a certain range of temperature.
They provide the effective spins seen in the Curie behavior below
the crossover and account for the entropy below $k_BT\sim J/5$
\cite{furu}. At a very
low energy scale these effective
spins will correlate among each other.
This picture has been confirmed by
numerical studies of finite size chains which show that the low-energy
dynamics is described by effective spins of varying sizes coupled by an
exchange interaction whose strength and sign depend
on the length and order of the segments \cite{long}.
This suggests that in order to study the low-temperature properties of the
random spin-$\frac{1}{2}$ chain, it is necessary to introduce an
effective Hamiltonian,
\begin{equation}
{\cal H}= \sum_i J_i {\bf S}_i \cdot {\bf S}_{i+1},
\label{eq:ham1}
\end{equation}
where the strength and sign of $J_i$ as well as the size of the spins
$S_i$ are random.
The great generality of the Hamiltonian (\ref{eq:ham1}) suggests that
apart from describing the low-energy behavior of
${\rm Sr}_3{\rm CuPt}_{1-x}{\rm Ir}_x {\rm O}_6$
it should be applicable
to a wider range of disordered 1D spin systems.

We investigate the low-temperature behavior of systems described by
(\ref{eq:ham1})
by means of a generalized version of the real-space
renormalization group (RG) method invented by Dasgupta and
Ma (DM) \cite{dasg} to study random bond AF $S=\frac{1}{2}$
Heisenberg chains (see also \cite{fish,hirsch}).
Let us define a link in the chain as two neighboring spins
and the coupling $J$ connecting them.
If a link were isolated from the rest of the chain, it would form
a local ground state of maximum ($J<0$) or minimum ($J>0$)
spin with an energy gap $\Delta$ to the first excited multiplet.
For a FM bond $\Delta = -J(S_L+S_R)$ while for an AF bond
$\Delta = J(|S_L-S_R|+1)$, where $S_L$ and $S_R$ are the left and right
spins of the link, respectively.
If we keep track of whether a link is ferromagnetic or antiferromagnetic,
there is a one-to-one correspondence between gaps and coupling constants
and we characterize from now on a link by \{$\Delta,S_L,S_R$\}.
Now we focus on the strongest link in the chain, defined as the link
with the largest gap $\Delta\equiv\Delta_0$.
If the distribution of gaps is broad, the gaps of the two neighboring
links, $\Delta_1$ and $\Delta_2$, are typically much smaller than
$\Delta_0$ so that the two spins $S_L$
and $S_R$, to a good approximation, lock into their local ground state.
Consequently we replace the link \{$\Delta_0,S_L,S_R$\} by a single
effective spin $\tilde{S}=|S_L\pm S_R|$ representing the local ground
state of minimum (AF) or maximum (FM) spin.
The weaker neighboring bonds are then taken into
account perturbatively in $\Delta_{1,2}/\Delta_0$, leading to an effective
interaction between the spins $S_1$, $\tilde{S}$, and $S_2$
(see inset in Fig.~\ref{fig:fixdist1D}).
Spin isotropy is conserved in this procedure, and to first order in
$\Delta_{1,2}/\Delta_0$ no next-nearest-neighbor interactions are generated.
The case of $\tilde{S}=0$ (corresponding to an AF link with $S_L=S_R$)
needs a special treatment, since to first order this would simply decouple
the left and right part of the  chain.
In this case we remove both $S_L$ and $S_R$ (which together form a
singlet) and make use of the fact that second-order processes introduce
a weak effective coupling between $S_1$ and $S_2$. This last step
corresponds to the DM treatment of the purely AF spin-$\frac{1}{2}$ model.

Integrating out the strongest link successively in the manner described
above preserves the form of the Hamiltoninan (\ref{eq:ham1}) but
changes the distribution of links and, in particular, lowers $\Delta_0$.
Thus the RG procedure generates a flow in the distributions of FM and AF
links,
\begin{equation}
P^{F,A} (\Delta_0; \Delta, S_L , S_R).
\label{eq:dist1}
\end{equation}
After having integrated out
links until the largest remaining gap in the chain is $\Delta_0$,
the chain is described by the Hamiltonian (\ref{eq:ham1}) with AF and
FM links distributed according to (\ref{eq:dist1}).
For low enough energies $\Delta_0$ we expect the distributions to flow
to fixed-point distributions where $P^F$ and $P^A$ exhibit scaling
behavior
\begin{equation}
P^{F,A}(\Delta_0;\Delta,S_L,S_R) =
\Delta_0^{-\gamma}Q^{F,A}(
{\textstyle \frac{\Delta}{\Delta_0^{\beta}},
\frac{S_L}{\Delta_0^{-\alpha}},\frac{S_R}{\Delta_0^{-\alpha}} }).
\label{eq:fixdist1a}
\end{equation}
In principle the exponents $\alpha$, $\beta$, and $\gamma$ in $P^F$
could differ from the ones in $P^A$.
However, this is very unlikely, because FM and AF links are
strongly interrelated with each other as we can see from the fact that
a large number of the spins belong to both an AF and a FM link and
that in the RG procedure an AF link can be converted into a FM link
and vice versa. Moreover, different exponents $\gamma$ would imply
that one of the probability distributions scales to zero, leaving us
with a purely FM or AF chain.
Indeed our numerical RG calculation shows that this is not the case.
In addition to the exponents in (\ref{eq:fixdist1a}) we are interested
in how length scales with $\Delta_0$.
Defining $n$ as the ratio of the original number of sites to
the number of effective spins, we expect
$n\sim\Delta_0^{-\delta}$.

The exponents in (\ref{eq:fixdist1a}) are not all independent.
{}From the normalization condition
$\sum_{S_L}\sum_{S_R}\int_0^{\Delta_0}\!\!(P^F+P^A) d\Delta = 1$
for any value of $\Delta_0$ follows $\gamma = \beta-2\alpha$.
Furthermore, if $Q^{F,A}$ describe fixed-point distributions
the average gap $\langle\Delta\rangle$ has to scale linearly in
$\Delta_0$, i.e., $\beta=1$.
Finally we argue that $\delta=2\alpha$.
An effective spin at energy scale $\Delta_0$ is built up by two more
`primitive' spins which in turn are built up by two even more
`primitive' spins etc.
The total spin of such a unit is just the sum of the participating
original spins, where each spin enters the sum with the same (opposite)
sign as its neighbor if the coupling is ferrromagnetic (antiferromagnetic).
Thus an effective spin made up of $n$ original spins is a sum of $n$
independent random variables and scales as $S^2\propto n$ (see below).
Thus, we find $n \propto S^2\propto
\Delta_0^{-2\alpha}$ and hence $\delta = 2\alpha$.

We have numerically performed the RG for five chains with
different starting distributions, including one purely AF distribution
and one distribution simulating the low-energy physics of
${\rm Sr}_3{\rm CuPt}_{0.8}{\rm Ir}_{0.2} {\rm O}_6$ \cite{footnote3}. In each
case
we keep the number
of sites fixed to $10^6$ by adding one link after each decimation and
iterate up to $2\cdot 10^7$ steps. Details of this calculation
will be presented elsewhere \cite{west1}. In all five chains the distribution
after appropriate scaling eventually approaches the
universal fixed-point distribution
(\ref{eq:fixdist1a}). Integrating $Q^F$ and $Q^A$ over two variables gives
the distribution of spins and the distribution of gaps plotted in
Fig.~\ref{fig:fixdist1D}.
The ratio of FM bonds to AF bonds is $0.59$ at the
fixed point and from the scaling of the averages
(see Fig.~\ref{fig:scaling}) the
exponents are determined to  be $\alpha = 0.22\pm0.01$,
$\beta = 1.00\pm0.005$ (in both $P^F$ and $P^A$) and
$\delta = 0.44\pm0.02$.
Thus the scaling form (\ref{eq:fixdist1a})
and the relations between the exponents are indeed verified in our
numerical calculations, and the fixed-point distributions become
\begin{equation}
P^{F,A} =
\frac{1}{\Delta_0^{1-2\alpha}}Q^{F,A}(\Delta/\Delta_0,
S_L\Delta_0^\alpha,S_R\Delta_0^{\alpha})
\label{eq:fixdist2a}
\end{equation}
and
\begin{equation}
n \sim \Delta_0^{-2\alpha} \ \ ; \ \ \alpha = 0.22\pm0.01 \ .
\label{eq:fixdist2b}
\end{equation}

The scaling forms (\ref{eq:fixdist2a}) and (\ref{eq:fixdist2b}) are the
key results from which we obtain the magnetic susceptibility $\chi$,
entropy $\sigma$, and specific heat $C$.
At finite temperature $T$ the RG flows will stop at $\Delta_0\sim k_BT$.
All pairs of spins in links with the gap larger than $\Delta_0$ are
frozen to form effective large spins according to the RG, while the
spins that have survived down to $\Delta_0$ are typically much more
weakly coupled and essentially free.
These free spins give a Curie contribution to the magnetic
susceptibility per unit length,
\begin{equation}
\frac{\chi}{L} =
\frac{\mu^2}{3k_B T}\frac{\langle {\bf S}^2\rangle}{n} =
\frac{c}{T}\frac{\Delta_0^{-2\alpha}}{\Delta_0^{-2\alpha}}
= \frac{c}{T}.
\label{eq:susc1}
\end{equation}
The Curie constant $c$ can be calculated in
the following way.
Our Hamiltonian (1) contains only nearest-neighbor coupling and,
therefore, does not give rise to any frustration.
In order to estimate $\langle{\bf S}^2\rangle$, the average size of
the effective spins, we consider an isolated
system with $n$ spins and search for the total spin quantum number
of its ground state.
{}From the multiplet of the ground state we choose the state with the largest
$z$-component, $S^z_{\rm tot}=S_{\rm tot}$ (total spin size).
This state contains the basis state  corresponding to the classical
ground state with all spin axes parallel to the $z$-axis.
The correlation of classical spins is given by
$S^z_{{\rm cl},i}S^z_{{\rm cl},j}=S_iS_j\prod^{i-1}_{k=j}{\rm sgn}(-J_k)$,
where $S_i$ is the size of the spin $i$.
The disorder averaged value of $ S^z_{\rm tot}$ can now be obtained within
the random walk picture as,
\begin{eqnarray}
\langle(S^z_{\rm tot})^2\rangle &=&
\sum^n_{i=0}(S^z_{{\rm cl},i})^2
 + 2\sum_{i>j}S^z_{{\rm cl},i} S^z_{{\rm cl},j} \cr
&=&
n\langle S^2_i \rangle
+2\langle S_i\rangle^2\sum^n_{i=1}\sum^{n-1}_{j=0}
         \prod^{i-1}_{k=j} {\rm sgn}(-J_k),
\end{eqnarray}
where $\langle S_i\rangle$ and $\langle S^2_i\rangle$ are
averages of the spin size of the original spin system (\ref{eq:ham1}).
Using the average $\langle{\rm sgn}(-J_l)\rangle=2q-1$
($q$: probability that a bond is ferromagnetic)
we find the Curie constant,
\begin{equation}
c= \frac{\mu^2}{3 k_B } \left[ \langle S^2_i \rangle
+\frac{2q-1}{1-q}\langle S_i \rangle^2 \right],
\label{eq:curieconst}
\end{equation}
in the limit $ T \to 0 $ ($ n \to \infty $). In particular, in the
case where all spins are $1/2 $, the low-temperature Curie constant
is $ c = (\mu^2/12k_B) q/(1-q) $, which
is in contrast to the high-temperature value $ c= \mu^2/4k_B $.

The entropy at zero external magnetic field follows from similar
arguments.
At a given temperature $T$ the spins that have not yet frozen into
large spins are essentially free and contribute to the entropy as
\begin{equation}
\frac{\sigma(T,H=0)}{L}\propto \frac{k_B\ln(2\langle S\rangle+1)}{n}
\propto \alpha k_B T^{2\alpha}|\ln T|,
\label{eq:entro1}
\end{equation}
which give the leading term of the specific heat as
\begin{equation}
\frac{C(T)}{L} \propto T^{2\alpha}|\ln T|.
\label{eq:heatcap1}
\end{equation}
In a finite magnetic field $ H $ and at finite temperature the RG flows are
interupted either by the thermal energy $k_BT$ or by the magnetic
energy $E_{\rm ZM}=\mu\langle S\rangle H$.
In the former case the magnetization is given by
(\ref{eq:susc1}) and the entropy by (\ref{eq:entro1}).
Otherwise, the magnetic field eventually drives the system away
from the fixed point of zero magnetic field into a state of aligned
effective spins where the entropy is zero and the
magnetization saturates.
A magnetic field $H$ starts to align the spins at an energy scale
$\Delta_0\sim\mu\langle S\rangle H$, i.e., $\Delta_0\sim H^{1/1+\alpha}$
and the saturated magnetization can be estimated to be
$M/L\sim\mu\langle S\rangle /n\sim H^{\alpha /1+\alpha}$.
The condition that the chain is not yet dominated by thermal
fluctuations thus becomes $k_BT<\Delta_0\sim H^{1/1+\alpha}$ and hence
\begin{equation}
\frac{M(T,H)}{L}\sim
\cases{
H^{\frac{\alpha}{1+\alpha}}, & $T^{1+\alpha}\ll bH$\cr
H/T, & $T^{1+\alpha}\gg bH,$\cr}
\label{eq:magn1}
\end{equation}
where $b$ is a non-universal dimensionful constant.
Similarly the entropy $\sigma(T,H)$ decreases rapidly towards zero at
$T^{1+\alpha}\sim\mu\langle S_i\rangle H$ when the magnetic field
starts to align the spins.

In the random bond AF spin-$\frac{1}{2}$ chain the DM RG treatment becomes
asymptotically exact \cite{fish} because the perturbation parameter
$\varepsilon=\Delta_t/\Delta_0$ approaches zero in the low-energy limit
($\Delta_t$: the typical gap in the chain).
This is not the case in our RG scheme
where $\varepsilon\sim0.2$ for the fixed-point distribution
(see Fig.~\ref{fig:fixdist1D}) and
higher order terms induce interactions other than
nearest-neighbor coupling.
However, such interactions between spins separated by
a distance $d$ have couplings of order $\varepsilon^d$.
This exponential dependence on $d$ is preserved in the RG
so that the application of an effective nearest-neighbor Hamiltonian
(\ref{eq:ham1}) is justified.
Furthermore, terms of the type $({\bf S}_i\cdot{\bf S}_{i+1})^m$
may appear in higher order perturbation.
Although they may change the local correlations, they would not affect the
relations between the exponents.
For this conclusion, it is important to remember that the effective spin
scales as $S \propto n^{1/2}$.
This scaling results from the random correlation of the spins within clusters
of length $n$ and can be understood immediately within a random walk
picture, which is quite robust against changes of details of the correlation.
A consequence of this is the relation between the exponents,
$\alpha=2\delta$ which reduces the number of free exponents to one, $\alpha$.
The only possible effect of higher order terms is to
modify slightly the exponent $\alpha$.
In addition, we emphasize that the Curie-like behavior of the magnetic
susceptibility (\ref{eq:susc1}) is independent of $\alpha$.

The low-energy physics of our system is very different from that of the
random AF spin-$\frac{1}{2}$ Heisenberg chain studied in
Refs.~\onlinecite{dasg,fish,hirsch}.
In the AF spin-$\frac{1}{2}$ chain the spins successively lock into singlets
which afterwards do not participate in the low-energy physics.
This {\it random singlet phase} \cite{bhat} is characterized by the formation
of singlets between spins that may be far apart from each other.
The coupling between two $S=\frac{1}{2}$ spins that have survived down to
some low-energy scale is mediated by virtually exciting all the singlets
in between the two spins into triplets.
This leads to a coupling that decreases exponentially with length,
$J=\Delta\propto\exp(-\sqrt{n})$.
This, together with the fact that the spin size
remains constant ($S=\frac{1}{2}$), leads to
susceptibility $\chi\propto T^{-1}|\ln T|^{-2}$ and entropy
$\sigma (T,H=0)\propto |\ln T|^{-2}$.
In our terminology this would correspond to $P^F=0$, $\alpha = 0$,
and $\delta = 0$ up to logarithmic corrections.
In contrast, our fixed point is characterized by large spins and a power
law relation between energy and length leading to (\ref{eq:fixdist2b}),
(\ref{eq:susc1}), (\ref{eq:entro1}), and (\ref{eq:heatcap1}).
It is clear that the two systems belong to different universality classes.
We emphasize here that the universality class of our model includes
a large variety of random spin models.
In particular, an AF spin chain with random spin sizes belongs to
this class.
(It is easy to see that such a system generates effective FM interaction
among effective spins as the RG iteration proceeds.)
Similarly, any spin chain with randomly distributed AF and FM bonds should
belong to this universality class \cite{footnote1}.
Hence the $S=\frac{1}{2}$ AF fixed point studied in
Refs.~\onlinecite{dasg,fish,hirsch} is unstable toward the presence of FM
couplings as well as toward randomness in the magnitude of the spins.
In either case we expect the system to flow to the fixed point discussed
in this paper.

An important candidate for this universality class is represented by
the compound Sr$_3$CuPt$_{1-x}$Ir$_x$O$_6$ mentioned earlier, where the
low-energy physics is described by the Heisenberg model (\ref{eq:ham1}).
This system shows three Curie regimes with different
Curie constants \cite{long}.
At high temperature it corresponds to independent
S$=1/2$ spins, the intermediate regime is governed by free effective
spins formed on the FM and AF segments, and the low-temperature regime
scales to our universality class.
The slow approach of $\langle n\rangle$ to the fixed point shown in
Fig.~\ref{fig:scaling} means that it will be difficult to observe the
fixed-point value of the specific heat
exponent whereas we find numerically that the limit of
$\chi$ given in (\ref{eq:susc1}) and (\ref{eq:curieconst}) should be more
readily observable.

We would like to thank T.~K.~Ng for helpful discussions and H.-C.~zur Loye
and T.~N.~Nguyen for stimulating our interest in the problem.
We are also grateful for the financial support by the Swedish Natural
Science Research Council (E.W.), by Swiss Nationalfonds
(M.S., No.~8220--037229), by the M.I.T. Science Partnership
Fund,
and by the NSF through the Material Research
Laboratory (DMR--90--22933).

\begin{figure}
\caption{(a) The fixed-point distributions of the spin sizes,
$P_{S}^{F,A}(S)\equiv\sum_{S_L}\int d\Delta Q^{F,A}$.
In the figure both distributions are normalized so that
$\sum_S P_{S}^{F,A}=1$.
(b) The fixed-point distributions for the gaps,
$P_{\Delta}^{F,A}(\Delta )\equiv\sum_{S_L,S_R}Q^{F,A}$. The distributions are
normalized as in (a).
Inset: Schematic picture of the decimation procedure.}
\label{fig:fixdist1D}
\end{figure}

\begin{figure}
\caption{Scaling of the averages which determine the exponents $\alpha$,
$\beta$, and $\delta$. The averages of the gap $\langle\Delta\rangle$ and spin
size $\langle S\rangle$ are
calculated from $P^F(\Delta_0;\Delta,S_L,S_R)$.
The corresponding averages for $P^A$ are almost identical. The initial
distribution simulates ${\rm Sr}_3{\rm CuPt}_{0.8}{\rm Ir}_{0.2}
{\rm O}_6$~\protect\cite{footnote3}.
We have found more rapid crossover to the fixed point for
other initial distributions,
{\it e.g.} the flat distribution $P^{F,A}={\rm const.}$, $-1<J<1$,
$S_{L,R}\in \{ 1/2,1,..,3\}$.}
\label{fig:scaling}
\end{figure}
\end{multicols}

\begin{references}
\bibitem{footnote2} Strictly speaking it is the occurence of Ir atoms
that is random so that FM bonds always appear in pairs.
\bibitem{footnote4} While a crossover between
different Curie regimes also appears in
the corresponding model with classical spins, the entropy of the
classical model is not sensitive to the disorder and it is
clearly necessary to study the quantum problem.
\bibitem{footnote3} To simulate
${\rm Sr}_3{\rm CuPt}_{0.8}{\rm Ir}_{0.2} {\rm O}_6$ we start with a
spin-$\frac{1}{2}$ chain with $20\%$ FM and $80\%$ AF bonds (all with strength
$J$)
and perform the segmentation described in \cite{long}. This gives the
distribution of couplings between the effective degrees of freedom
which we take as initial distribution in
the DM RG.
\bibitem{footnote1}
 Obviously random FM systems of any type are excluded from this class.
\end{references}
\end{document}